\documentclass[prd,twocolumn,showpacs,floatfix]{revtex4}
\usepackage{graphicx}
\usepackage{amsfonts}

\def\cA{{\cal A}}

\def\cC{{\cal C}}
\def\cD{{\cal D}}

\def\cL{{\cal L}}
\def\cM{{\cal M}}
\def\qq{{\overline q}q}
\def\be{\begin{equation}}
\def\ee{\end{equation}}
\newcommand{\GS}[1]{#1\!\!\!\!\!\not~}

\newcommand{\tr}{\mbox{tr}}
\def\Ds {{D \hspace{-6.4pt} \slash}\;}

\begin{document}
\title{Extracting the $\qq$ condensate for light quarks\\[0.5mm]
 beyond the chiral limit in models of QCD}
\author{R. Williams$^{1}$}
\author{C. S. Fischer$^{1,2}$}
\author{M. R. Pennington$^{1}$}
\affiliation{$^{1}$Institute for Particle Physics Phenomenology, Durham University, Durham  DH1 3LE, UK}
\affiliation{$^{2}$ Institut f\"ur Kernphysik, Darmstadt University of Technology, Schlossgartenstra\ss e 9, 64289 Darmstadt, Germany}

\begin{abstract}
	It has recently been suggested~\cite{Chang:2006bm} that a reliable and unambiguous definition of the non-perturbative massive quark condensate could be provided by considering a non positive-definite class of solutions to the Schwinger Dyson Equation for the quark propagator. In this paper we show that this definition is incomplete without considering a third class of solutions. Indeed, studying these three classes reveals a degeneracy of possible condensate definitions
leading to a whole range of values. However, we show that the {\it physical} condensate may in fact be extracted by simple fitting to the Operator Product Expansion, a procedure which is stabilised by considering the three classes of solution together. We find that for current quark masses in the range from zero to 25 MeV or so (defined at a scale of 2 GeV in the $\overline{MS}$ scheme), the dynamically generated condensate increases from the chiral limit in a wide range of phenomenologically successful models of the confining QCD interaction. Lastly, the role of a fourth class of noded solutions is briefly discussed.
\end{abstract}
\pacs{12.38.-t, 11.30.Rd, 12.38.Aw, 12.38.Lg}

\maketitle
\section{Introduction}
A remarkable aspect of strong coupling field theory is the possibility that masses can be largely, or even entirely, created by interactions. It is by this mechanism that
%Dynamical chiral symmetry breaking (DCSB) is the generation of 
a fermion mass gap is generated for the light flavours in QCD, leading to dynamical chiral symmetry breaking (DCSB). Indeed, the mass gap generated far exceeds the scale of the current quark masses present in the Lagrangian. The strong coupling induces long range ${\overline  q}q$ correlations that polarise the vacuum. It is the scale of these condensates that determines the mass generated, and this gap persists even in the chiral limit. As long known this underpins much of QCD phenomenology. In this paper we investigate this mass generation using the Schwinger-Dyson equations. The aim is to extract the behaviour of the $\qq$ condensate beyond the limit of zero quark mass.

The interest in the value of such a condensate arises in the context of QCD sum-rules. There the Operator Product Expansion (OPE) is used to approximate the short distance behaviour of QCD. In studying currents like that of $\,{\overline q_i} \gamma^{\mu} (\gamma_5) q_j$, with $q_i=s$ and $q_j=u,d$, the vacuum expectation values of ${\overline  u}u$, ${\overline  d}d$ and ${\overline  s}s$ operators naturally arise~\cite{Jamin:2002ev,Jamin:2001fw,Jamin:2006tj}. In the chiral limit, the value of the $\qq$ condensate for the $u$ and $d$ quarks is well determined to be $-(235 \pm 15 \ {\rm MeV})^3$ by experiment --- in particular from the low energy behaviour of $\pi\pi$ scattering~\cite{colangelo}. However, in the OPE it is the value of the condensates away from the chiral limit that actually enters. Since the current masses of the $u$ and $d$ quarks are only a few MeV, the resulting condensate is expected to be close to its value in the chiral limit, but how close? For the first 20 years of QCD sum-rules
their accuracy was never sufficient for it to matter whether this
difference was a few percent, 10\% or even 20\% effect.
This equally applied to the estimate by Shifman, Vainshtein and Zakharov~\cite{Shifman:1978bx,Shifman:1978by} that
the ${\overline s}s$ condensate was $(0.8\,\pm\,0.3)$ of the ${\overline u}u$ and ${\overline d}d$ values. It is the greater precision brought about  by the studies of Refs.~\cite{Dominguez,Maltman:2001jx,Maltman:2002sb,
Jamin:2006tj}, for instance, that motivate the need to learn about how the ${\overline q}q$ condensate depends on the current quark mass. Indeed, Dominguez, Ramlakan and Schilcher~\cite{Dominguez} compute the ${\overline s}s$ condensate to be just $(0.5\,\pm\,0.1)$ times that for ${\overline u}u$ and ${\overline d}d$. In the light of a better understanding~\cite{Roberts:1994dr,Alkofer:2000wg,Maris:2003vk,Fischer:2006ub}  of strong coupling QCD how robust is this? First results were already presented in Ref.~\cite{williamspl}.

For light quarks, $u$, $d$ and $s$, studying the 
Schwinger-Dyson equation for the fermion propagator in the continuum is essential, until computation with large lattice volumes  become feasible.
Since the continuum Schwinger-Dyson equations can be solved for any value of the quark mass, they also provide a natural way to bridge the gap between lattice data at larger masses and the chiral limit of phenomenological importance.
Our primary focus is, of course, on QCD, but we shall draw on the NJL model where necessary.
\section{Schwinger-Dyson Equations}
\begin{figure}[b]
\vspace{1mm}
 \begin{center}  
       \includegraphics*[width=0.95\columnwidth]{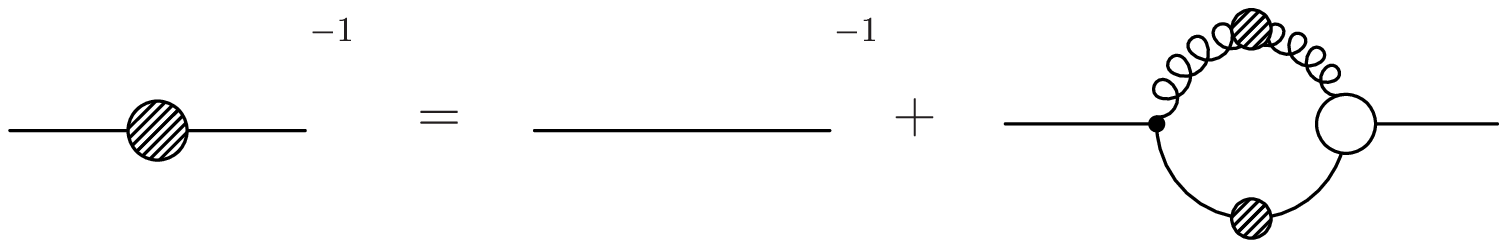}
	\caption{Schwinger-Dyson equation for the quark propagator}\label{quark:sde}
\end{center}  
  \end{figure} 
Solving the Schwinger-Dyson Equations for QCD is an entirely 
non-trivial process, see 
\cite{Roberts:1994dr,Alkofer:2000wg,Maris:2003vk,Fischer:2006ub} 
for reviews. They comprise an infinite tower of coupled 
integral equations, that can be solved analytically only for 
specific kinematical situations, for instance in the far infrared 
and in the ultraviolet~\cite{Lerche:2002ep,Alkofer:2004it,Alkofer:2006gz,Fischer:2006vf}.
 In general,
the solutions have to be found numerically and then only after 
some form of truncation has been applied. To this end we require 
some suitable ansatz for the three-point functions in order to 
allow us to solve self-consistently for propagators. 
In the following sections we will 
use a number of truncation schemes that have been tested elsewhere
\cite{Maris:1997hd,Fischer:2003rp,Fischer:2005nf}, within which 
we may solve for the fermion propagator. 
Our chief aim is to calculate the mass function of the quark propagator for a range of current masses. Our starting point is the renormalized Schwinger-Dyson equation for the quark propagator as depicted in Fig.~\ref{quark:sde}:
\begin{eqnarray}\label{cond:eqn:quarksde1}	% The Quark Schwinger-Dyson Equation.
\vspace{2mm}
	S_F^{-1}(p) &=&  Z_2\left[ S^{(0)}(p) \right]^{-1}-C_F \frac{\tilde{Z}_1\,Z_2}{\tilde{Z}_3}\frac{g^2}{\left( 2\pi \right)^4}\int d^4k \nonumber\\[2mm]
	&&\hspace{35pt}\times\gamma_\mu S_F(k)\Gamma_\nu(k,p)D_{\mu\nu}(p-k)\;.
\vspace{2mm}
\end{eqnarray}
In the Landau gauge we can choose~\cite{Taylor} $\tilde{Z}_1 = 1$. The inverse propagator $S_F^{-1}(p)$ is specified by two scalar functions $\cA$ and $\cM$:
\begin{equation}\label{fermion:eqn:AB}
	S_F^{-1}(p)\,=\,\cA(p^2)\left(\GS{p} +\cM(p^2)\right)\; .
\end{equation}
While $\cA$ is also a function of the renormalisation point $\mu$ and so strictly $\cA(p^2,\mu^2)$, the quark
mass function $\cM(p^2)$ is renormalisation group invariant.
Projecting  out these two functions from Eq.~(\ref{cond:eqn:quarksde1}), we have two coupled equations to solve.

The operator product for the mass function has an expansion at large momenta given symbolically by:
\begin{equation}\label{ope-simple}
\vspace{2.5mm}
\cM(p^2)\,\simeq\,{\overline m}(p^2)\;+\;\frac{\rm const}{p^2}\,\langle {\overline q}q (p^2)\rangle\;+\cdots\; ,
\vspace{1mm}
\end{equation}
where the first term corresponds to the explicit mass in the Lagrangian, and the second to the lowest dimension vacuum condensate.  For now we just show the momentum dependence given by the canonical dimensions, and leave for later the implications of the anomalous dimensions of QCD. Having computed the mass function using the Schwinger-Dyson equations, the essential problem is how to separate these two terms in Eq.~(\ref{ope-simple}) with any accuracy if ${\overline m}$ is non-zero. We note that the mass function for a physically meaningful solution 
is expected to be positive definite.

 In the chiral limit, there exist three solutions for the mass function $\cM(p^2)$. These correspond to the Wigner mode (the only solution accessible to perturbation theory), and two non-perturbative solutions of equal magnitude generated by the dynamical breaking of chiral symmetry. These we denote by:
\begin{equation}\label{cond:eqn:sols} 
\vspace{1mm}
	\cM(p^2) = 	\left\{ \begin{array}{l} 
				M^W(p^2) = 0\;\; \\ [-0.5mm]
                               \\[-0.5mm]
			      M^\pm(p^2) = \pm M^0(p^2) 
			\end{array} \right. \;.
\vspace{1mm}
\end{equation}
%Usually one only concentrates on the positive-definite solution $M^+(p^2)$, and this continues to define the physically meaningful solution even when a non-zero quark mass is employed. However,
Such multiple solutions have been found in the context of QED$_4$ by Hawes {\it et al.}~\cite{Hawes}.
One can ask whether analogous solutions exist in QCD as we move away from the chiral limit and what relevance they hold. Indeed, in a recent paper~\cite{Chang:2006bm} it was suggested that one could make an unambiguous definition of the massive quark condensate by taking a particular combination of these solutions. The existence of these  is restricted to the domain:
\begin{equation}\label{domain}
	\cD = \left\{ m : 0 \le m \le m_{cr}\right\}\, ,
\end{equation}
where only the positive-definite solution exists beyond $m_{cr}$. Chang {\it et al.}~\cite{Chang:2006bm} found that inside the critical domain, the solutions for both $M^+(p^2)$ and $M^-(p^2)$ exhibit the same running current-quark mass in the ultraviolet.
In terms of Eq.~(\ref{ope-simple}), this means both solutions have the same ${\overline m}$ term.  Noting that the $M^-(p^2)$ solution had a condensate of opposite sign, they proposed a definition of the massive quark condensate given by:
\begin{eqnarray}\label{cond:eqn:cond-roberts}
\vspace{1.5mm}
	{\overline \sigma}(m(\mu))&=&\lim_{\Lambda\rightarrow\infty}\,Z_4(\mu, \Lambda)\, N_c \,\tr_D \int_k^\Lambda\,\frac{d^4k}{(2\pi)^4}\nonumber\\
	&&\hspace{62pt}\times\; \frac{1}{2}\left[ S_+\,-\,S_- \right]\;.
\vspace{1.5mm}
\end{eqnarray}
where $\mu$ denotes the renormalisation point.
Since both propagators $S_\pm$ share the same asymptotic behaviour
$\cM(p^2) \rightarrow {\overline m}(p^2)$ for large momenta, potential
divergences in the integral cancel and the expression 
Eq.~(\ref{cond:eqn:cond-roberts}) is well defined. The resulting condensate
${\overline\sigma}(m(\mu))$ will be equal to the one for the physical 
$M^+$ solution provided $S_\pm$ have condensates of equal magnitude  
away from the chiral limit. However, we will show that this assumption is
not correct. Morever, we find here that there is in fact an analogous 
solution $S_W$ to the Wigner mode of the chiral limit. (This has also been noted
 in the 
revised version of~\cite{Chang:2006bm}.) This solution has the same ultraviolet 
behaviour of a running current quark mass, {\it i.e.} all three, $M_\pm$ and 
$M_W$, have a common ${\overline m}(p^2)$ in their OPE, see Eq.~(\ref{ope-simple}). Consequently, the combination
\begin{equation}\label{fermion:eqn:beta}
S(\beta)\;=\;(2 - \beta)\, S_+\,-\,\beta S_-\,+\,2(\beta - 1)\,S_W
\end{equation}
has its asymptotics controlled by the second term of the OPE, Eq.~(2), for any $\beta$.
Thus, we can extend the definition of Chang {\it et al.}~\cite{Chang:2006bm} to a family of condensates parametrised by $\beta$:
\begin{eqnarray}\label{cond:eqn:cond-beta}
\vspace{1.5mm}
	{\overline \sigma}(m(\mu),\beta)\; =  \lim_{\Lambda\rightarrow\infty} Z_4(\mu, \Lambda)\, N_c\, \tr_D \int_k^\Lambda \frac{d^4k}{(2\pi)^4} \frac{1}{2}\;S(\beta)\; ,
\vspace{1.5mm}
\end{eqnarray}
where all $S_-$, $S_+$ and $S_W$ are dependent upon the momentum $k^2$, the renormalisation point $\mu$ and the quark mass $m(\mu)$.
The choice $\beta=1$ corresponds to the definition of Eq.~(\ref{cond:eqn:cond-roberts}), with $\beta=0$ and $\beta=2$ corresponding to two other natural choices. In fact $\beta$ could take any value from $-\infty$ to $+\infty$ and so we can have a whole range of values for the massive quark condensate: all agreeing in the chiral limit. 
Consequently, the definition proposed by Chang {\it et al.} is far from unique, and does not provide a value for the condensate that corresponds to the physical $M^+$ solution of interest. Eq.(\ref{cond:eqn:cond-roberts}) merely defines the value for the
difference of the condensates for $M^+$ and $M^-$. However, we will show that when combined with the OPE the 3 solutions will pick out a precise {\it physical} definition of this condensate.  Before investigating this in the context of 
%the model employed within~\cite{Chang:2006bm,} 
models of the QCD interaction, we shall draw analogy with the NJL model, within which a natural definition of the massive quark condensate already exists.
%\newpage

\section{The Nambu-Jona-Lasinio Model}\label{cond:sect:njl}
Because of the complexity of QCD, it is often prudent to examine simpler systems exhibiting similar characteristics first. One such example is the Nambu-Jona-Lasinio (NJL) model. Though originally formulated to describe nucleon interactions in the pre-QCD era, the model can be 
reinterpreted by regarding the nucleons as quarks~\cite{Eguchi:1976iz,Kikkawa:1976fe}. It can then be used to study bound states and so determine basic phenomenological quantities of meson interactions.
%\noindent
%It is believed that one of the defining features of QCD is the dynamical chiral symmetry breaking (DCSB).
 The NJL model shares the same symmetry structure as QCD, and is dominated by DCSB effects at low energies too. The Lagrangian for the NJL model with just two flavours of quarks with degenerate mass $m_0=m_u=m_d$ is:
\begin{equation}
	\cL_{NJL} = {\overline \psi}(x)\left( i\;\Ds-m_0 \right)\psi(x) +\cL_{int}\,.
\end{equation}
%
%where the subscript on the fermion fields denotes the spacetime dependence of the fields.
The interactions are given by a four-fermion contact term:
\begin{eqnarray}
	\cL_{int} &=&  \frac{G_\pi}{2}\left[ \left( {\overline \psi}(x)\psi(x) \right)^2+\left( {\overline \psi}(x) i\gamma_5\tau^a\psi(x) \right)^2 \right]
	\,, %\nonumber\\	&+& \frac{G_2}{2}\left[ \left( {\overline \psi}_x\gamma_\mu\tau^\alpha\psi_x \right)^2+\left( {\overline \psi}_xi\gamma_\mu\gamma_5\tau^a\psi_x \right)^2 \right]\,,
\end{eqnarray}	% check this equation is true!
%The four interaction terms are the scalar, pseudoscalar, vector and axial interactions respectively. 
with the two terms corresponding to the scalar and pseudoscalar channels respectively. From this can be derived the so-called gap equation:
\begin{equation}\label{cond:eqn:gap1}% Klevansky --- exchange interaction is the 1/2 factor. We will ignore this.
	m\,=\,m_0\,+\,i\;G_\pi N_c N_f \int\frac{d^4p}{\left( 2\pi \right)^4\, }\,\tr_D S(p)\;,
\end{equation} 
where the trace is over spinor indices in $D$-dimensions. At this point one has a choice of how to regulate the integrals. Since we are dealing with a non-renormalisable effective theory, our results will depend upon the cut-off used. One may introduce a non-covariant cut-off in the Euclidean 3-momentum, or employ a variety of covariant regularisation schemes such as a four-momentum cut-off, proper time or Pauli-Villars. The four-momentum cut-off is most closely related to the scheme we will employ in the later sections, so we choose this method. By inserting the form of the propagator into Eq.~(\ref{cond:eqn:gap1}) we arrive at
\begin{figure}[t]
 \begin{center}  
      \includegraphics*[width=0.95\columnwidth]{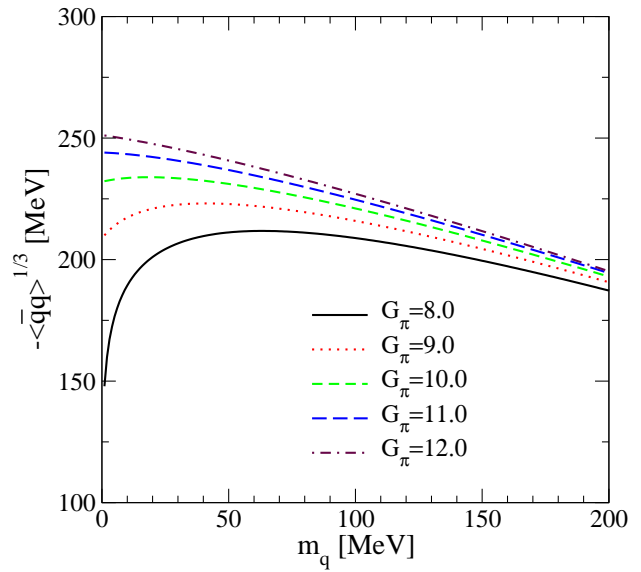}
	\caption{The non-perturbative condensate as a function of the current quark mass for a selection of NJL couplings $G_\pi$.}\label{cond:njl-p34}
\end{center}  
%\vspace{-3mm}
  \end{figure}
\begin{equation}\label{cond:eqn:gap2} % Benhaddou
	m\,=\,m_0\,+\,m\, G_\pi N_c\, 8 i\int^\Lambda \frac{d^4k}{(2\pi)^4}\frac{1}{k^2-m^2}\;,
\end{equation}
where we have introduced  a cut-off $\Lambda$ on the 4-momentum. Rotating to Euclidean space we obtain
\begin{equation}\label{cond:eqn:gap3} % Benhaddou
	m\,=\,m_0\,+\,\frac{G_\pi N_c m}{2\pi^2}\left[ \Lambda^2-m^2\log\left( 1+\frac{\Lambda^2}{m^2} \right) \right]\;.
\end{equation}
The coupling $G_{\pi}$ and cut-off, $\Lambda$, are fixed by fitting to experimental data. To do this we must calculate the order parameter associated with the breaking of chiral symmetry. Thus the chiral condensate is given by:
\begin{equation}\label{cond:eqn:pcond}
	\left< {\overline q}q \right> = -i\;N_c\int \frac{d^4k}{\left( 2\pi \right)^4}\;\tr_D S_F(k; m)\;,
\end{equation}
and for non-zero current masses we define:
\begin{equation}\label{cond:eqn:npcond}
\vspace{1mm}
	\left< {\overline q}q \right> = -i\;N_c\int \frac{d^4k}{\left( 2\pi \right)^4}\;\tr_D\left[S_F(k; m)-S_F(k; m_0)\right]\;.
\vspace{1mm}
\end{equation}
Working with explicitly massless quarks and employing a covariant cut-off in the Euclidean 4-momentum, the pion-decay constant is given by~\cite{Klevansky:1992qe}:
\begin{equation}\label{cond:eqn:fpi2}
\vspace{1mm}
	f_\pi^2\,=\,N_c\, m^2\left[ \log\left( 1+\frac{\Lambda^2}{m^2} \right) -\frac{\Lambda^2}{m^2+\Lambda^2} \right]
%\dot{,}
\; .
\vspace{1mm}
\end{equation}
By solving Eq.~(\ref{cond:eqn:pcond}) and Eq.~(\ref{cond:eqn:fpi2}) simultaneously and demanding that $f_\pi = 93$ MeV and $-\left<{\overline q}q\right>^{1/3}=235$ MeV, we obtain $\Lambda = 0.908$ GeV and $m = 265$ MeV. These can be substituted into the mass gap equation of Eq.~(\ref{cond:eqn:gap3}) and thus we can solve for the coupling, finding $G_\pi = 10.2$.

%\noindent
Our interest is to investigate the condensate's dependence on the current quark mass. In Fig.~\ref{cond:njl-p34} we see that the behaviour of the condensate depends upon the chosen coupling. Indeed, it either increases to a maximum then decreases, or is monotonically decreasing for larger couplings as the quark mass increases. Moreover, one finds that the condensate can become vanishingly small for sufficiently large masses ($\sim\,600$ MeV, not shown in Fig.~\ref{cond:njl-p34}). %Fig.~\ref{cond:njl-p5} shows the behaviour of the condensate for our favoured parameter set.
% with $G_\pi = 7.63$ and $\Lambda = 1.015$ GeV.

\begin{figure}[t]
 \begin{center}  
      \includegraphics*[width=0.95\columnwidth]{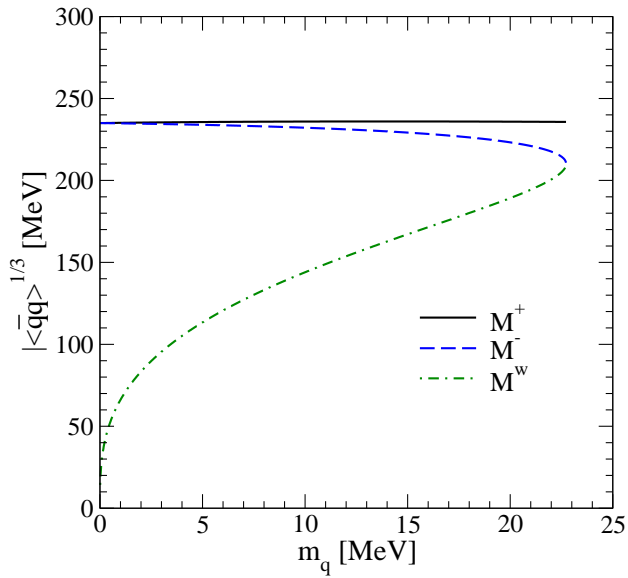}
	\caption{The quark condensate within the NJL model for the three solutions $M^{\pm,W}$ as functions of quark mass.}\label{cond:njl-cond3}
    \end{center}
\vspace{-3mm}
\end{figure}

Now, within the NJL model we also have solutions corresponding to $M^+$, $M^-$ and $M^W$ away from the chiral limit within some domain. The extent of the
 domain depends on the parameters of the model. With the favoured choice, $m_{cr} \simeq 15$ MeV in Eq.~(\ref{domain}). One may then use Eq.~(\ref{cond:eqn:npcond}) to calculate the massive quark condensate for these solutions individually, the result of which is shown in Fig.~\ref{cond:njl-cond3}. 
%\newpage
\begin{figure}[b]
%\vspace{3mm}
 \begin{center}
      \includegraphics*[width=0.95\columnwidth]{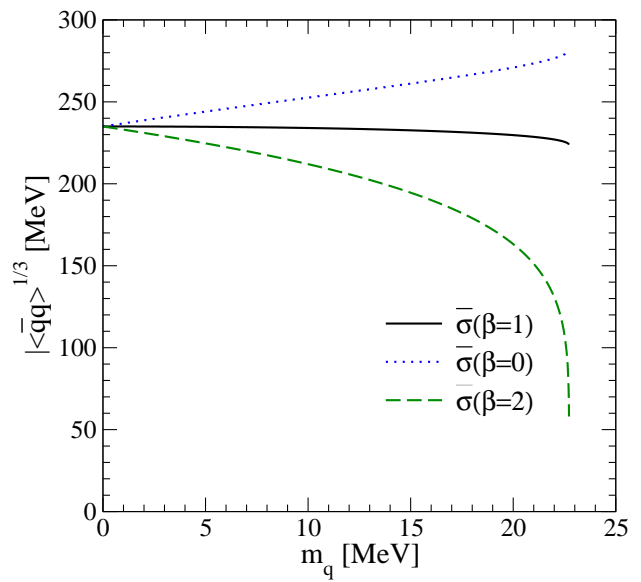}
	\caption{The condensate defined by ${\overline \sigma}(\beta)$ within the NJL model for values $\beta=\left\{0,1,2\right\}$ as a function of quark mass.}\label{cond:njl-comb3}
    \end{center}
%\vspace{-3mm}
\end{figure}
Furthermore, using the definition, Eq.~(\ref{cond:eqn:cond-beta}), gives a family of condensates, ${\overline \sigma}(m(\mu),\beta)\ $, for which 
%%%%% viz:
%\begin{eqnarray}\label{cond:eqn:njl-beta}
%	{\overline \sigma}(\beta)\, =\,  -i\,N_c\int\frac{d^4k}{\left( 2\pi \right)^4}\, \tr_D\, \frac{1}{2}\bigg[ (2-\beta)\,S_+-\beta S_-\nonumber\\
%	+2(\beta-1)\,S_W \bigg]\;.\nonumber\\[-7.3mm]
%\end{eqnarray}
the results are shown in Fig.~\ref{cond:njl-comb3} for $\beta=\{0,1,2\}$. What is clearly evident is that the condensates for the three mass functions $M^\pm$, $M^W$ are not equal. Drawing analogy with the Schwinger-Dyson solutions for QCD, one should therefore not expect the condensate of $M^\pm$ to be equal in magnitude as well as opposite in sign, as assumed by Chang {\it et al.}~\cite{Chang:2006bm}. We show this to be the case in the next section.

\section{Phenomenological Model of QCD interaction}\label{cond:sect:maris}
We now turn to QCD.
Rather than solving for the ghost and gluon system, one may employ some suitable ansatz for the coupling which has sufficient integrated strength in the infrared so as to achieve dynamical mass generation. There have been many suggestions in the literature~\cite{Maris:1997hd,Maris:1998hc} which have been extensively studied. Following the lead of Maris {\it et al.}~\cite{Maris:1997hd,Maris:1998hc}, we will employ an ansatz for $g^2D_{\mu\nu}(p-k)$ which has been shown to be consistent with studies of bound state mesons. We will consider other modellings in a later section. Since this simple model assumes a rainbow vertex truncation, the solutions are not multiplicatively renormalisable and so depend on the chosen renormalisation point. For comparison with earlier studies we take this to be $\mu=19$ GeV. We will scale results by one loop running to 2 GeV in the modified momentum subtraction scheme relevant to the Maris-Tandy model. Thus we use:
\begin{eqnarray}
%\vspace{1mm}
	\frac{g^2}{4\pi}\frac{Z_2}{\tilde{Z}_3}D_{\mu\nu}(q)\rightarrow \alpha\left(q^2\right) D^{(0)}_{\mu\nu}(q)
%\vspace{1mm}
\end{eqnarray}
where the coupling is described by:
\begin{eqnarray}
%\vspace{1mm}
	\alpha\left( q^2 \right) &=& \frac{\pi}{\omega^6}D q^4 \exp(-q^2/\omega^2)\nonumber\\[2.mm]
	&+&\frac{2\pi \gamma_m}{\log\left( \tau+\left(1+q^2/\Lambda_{QCD}^2 \right)^2\right)}\nonumber\\[0.mm]
	&&\times \left[ 1-\exp\left(-q^2/\left[ 4m_t^2 \right]\right) \right]\;,
%\vspace{1mm}
\end{eqnarray}
with 
%$m_t=0.5$ GeV, $\tau=\textrm{e}^2-1$, $\gamma_m= 11/33$ and $\Lambda_{QCD}=0.234$ GeV. 
\begin{eqnarray}
\nonumber m_t&=& 0.5\;{\rm GeV}\qquad\qquad,\qquad\tau\;=\;\textrm{e}^2-1\qquad\;,\\
\nonumber \gamma_m&=&12/(33-2N_f)\,\quad,\quad\Lambda_{QCD}\;=\;0.234\,{\rm GeV}\; .
\end{eqnarray}
Note that we work in the $N_f=0$ limit first since in Sect.~\ref{cond:sect:sophistry} we will  investigate the mass dependence of the condensate using a model derived from quenched lattice data~\cite{Fischer:2005nf}. The precise value of $\Lambda_{QCD}$ is irrelevent for our current study, and we choose the parameter set $\omega=0.4$ GeV, $D=0.933$ GeV$^2$ in the range considered by Ref.~\cite{Alkofer:2002bp}.

%\newpage
Solutions are obtained by solving the coupled system of fermion equations for $\cA$ and $\cM$ of Eq.~(\ref{fermion:eqn:AB}), which we may write symbolically as:
\begin{eqnarray}
	\cA(p^2,\mu)&=&Z_2(\mu,\Lambda) - \Sigma_D\left(p,\Lambda\right)\; ,\nonumber\\[-1mm]
&& \\[-1.mm]
	\cM(p^2)\cA(p^2,\mu)&=&Z_2(\mu,\Lambda) Z_m m_R(\mu) + \Sigma_S\left(p,\Lambda\right)\;.\nonumber
\end{eqnarray}
The $\Sigma_S$ and $\Sigma_D$ correspond to the 
%dirac odd and even projects 
scalar and spinor projections of the integral in Eq.~(\ref{cond:eqn:quarksde1}).
%\newpage
\begin{figure}[t]
\vspace{1mm}
 \begin{center}  
       \includegraphics*[width=\columnwidth]{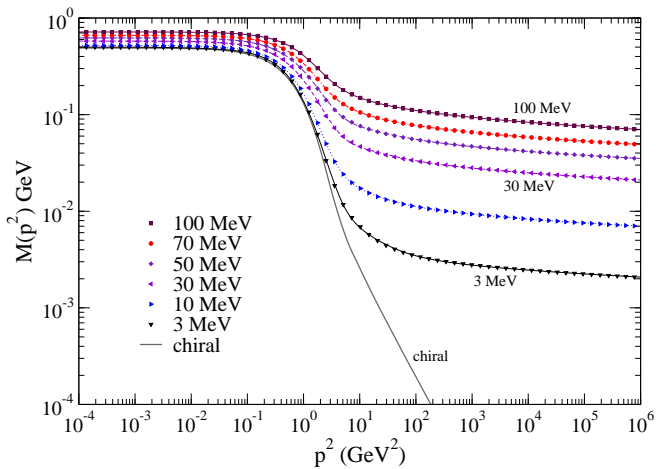}
	\caption{Euclidean mass functions for different current masses, specified at $\mu = 19$ GeV as labelled. The plot illustrates how on a log-log plot the behaviour dramatically changes between a current mass of 0 and 3~MeV.
These results are essentially the same as found by Maris and Roberts~\protect{\cite{Maris:1998hc}}.}\label{masses}
\end{center}  
\vspace{-4.mm}
  \end{figure} 
For massive quarks we obtain the solution $M^+$ by eliminating the renormalisation factors $Z_2$, $Z_m$ via:
\begin{eqnarray}
	Z_2(\mu,\Lambda)&=& 1+\Sigma_D\left( \mu,\Lambda \right)\; ,\nonumber\\[-2.mm]
  && \\[-2.mm]
	Z_m(\mu,\Lambda)&=& \frac{1}{Z_2(\mu,\Lambda)}-\frac{\Sigma_S\left(\mu,\Lambda \right)}{Z_2(\mu,\Lambda) m_R\left(\mu\right)}\;.\nonumber
\end{eqnarray}
The resulting momentum dependence for different values of $m_R$ are shown in Fig.~\ref{masses}. Our purpose is to define the value of the $\qq$ condensate for each of these.

\begin{figure}[t]
%\vspace{3mm}
	\centering
	%{\includegraphics*[width=0.90\columnwidth]{figures/three-mass-sol-16}}
	{\includegraphics*[width=0.98\columnwidth]{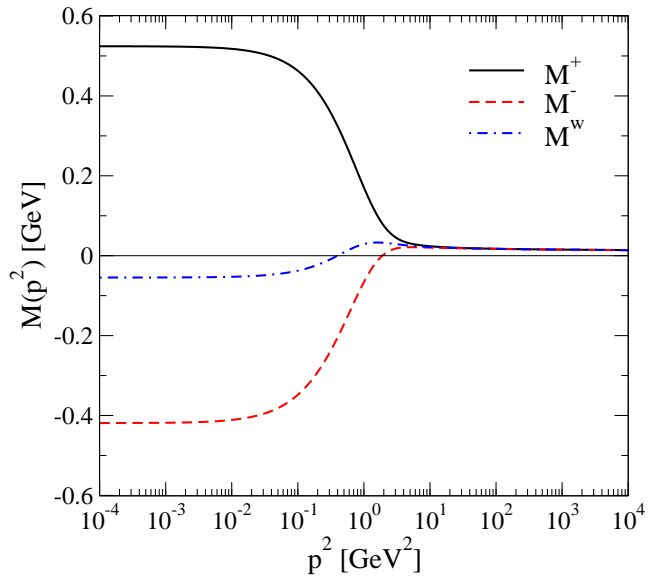}}
	\caption{Momentum dependence of the three solutions $M^\pm(p^2)$ and $M^W(p^2)$ for a quark mass $m(\mu)$=16 MeV, $\mu=19$ GeV.}\label{cond:fig:quark-sols}
%\vspace{2mm}
\end{figure}

At very large momenta the tail of the mass function is described by the operator product expansion~Eq.~(\ref{ope-simple}).
Crucially, for a given $m_R$ we obtain the $M^-$ and $M^W$ solutions by inserting the same $Z_2$ and $Z_m$ found for the $M^+$ solution. This ensures that differences in the dynamics of the three systems do not influence the ultraviolet running of the current-quark mass in the context of the subtractive renormalisation scheme used  by Maris and Tandy. The iteration process is performed using
Newton's method. For the solution $M^W$ this is mandatory, since it corresponds to a local maximum of the effective action and is therefore not accessible using the conventional fixed point iteration scheme.
A representative example of the solutions is shown in Fig.~\ref{cond:fig:quark-sols}.

 The value of the critical mass is model-dependent, and is summarised in Table~\ref{table:mariscrit}. 
Chang {\it et al.}~\cite{Chang:2006bm}  imbue this critical mass  with some significance for the dynamics of QCD. However, criticality does not feature in the {\it physical} solution $M^+$, which exists for all values of $m_q$. It only occurs in the $M^-$ and $M^W$ solutions, which appear in a
 strongly model-dependent region. Consequently, we find little evidence of 
criticality being
important to the mass generation in QCD. We will comment again on this when we consider more sophisticated vertex structures in a later section.  

\begin{table}[b]
\vspace{4mm}
\begin{center}
	\begin{tabular}{|c|c|cccr|}
\hline
	~$N_f$~ & $\omega$  & ~~0.3~ & ~0.4~ & ~0.5~ &~[GeV]~~ \\[1.mm]\hline\hline
&&&&&\\[-4.mm]
0 & \rule{0ex}{2.5ex} ~$m_{cr}(\mu=\,19$~GeV)~ & 38 & 34 & 16 &[MeV]~~\\[1.mm]
0 & \rule{0ex}{2.5ex} ~$m_{cr}(\mu=\,2$~~GeV)~ & 49 & 44 & 21&[MeV]~~\\[1.mm]\hline
&&&&&\\[-4.mm]
4 & \rule{0ex}{2.5ex} ~$m_{cr}(\mu=\,19$~GeV)~ & 35 & 31 & 16&[MeV]~~\\[1.mm]
4 & \rule{0ex}{2.5ex} ~$m_{cr}(\mu=\,2$~~GeV)~ & 49 & 44 & 23&[MeV]~~\\[1.mm]
\hline
\end{tabular}
\caption{How the critical mass that defines the domain of solutions Eq.~(\ref{domain}) depends on the number of quark flavours, $N_f$, on the gluon range parameter $\omega$, in the Maris-Tandy model. This critical mass is listed at two different renormalization scales, 19 GeV of Ref.~\cite{Chang:2006bm} and 2 GeV for ease of comparison with other works in a momentum subtraction scheme.  }\label{table:mariscrit}
%\end{equation}
\end{center}
\end{table}

\begin{figure}[t]
%\vspace{2mm}
	\centering
	%{\includegraphics*[width=0.90\columnwidth]{figures/condensate-beta-parameter-at1}}
	{\includegraphics*[width=0.95\columnwidth]{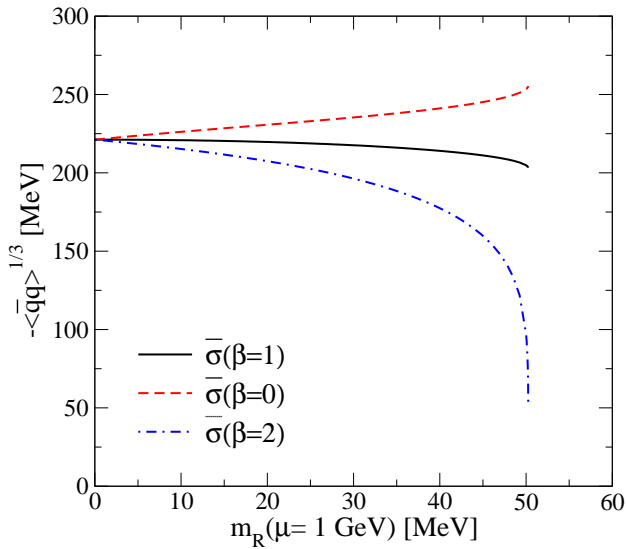}}
	\caption{Renormalisation point independent quark condensate as a function of $m_q$ as defined by Eq.~(\ref{cond:eqn:cond-beta}) for three values of $\beta$, showing how they are quite different despite the solutions having the same running current-mass and being equal in the chiral limit.}\label{cond:fig:quark-beta}
\end{figure}

The definition $S(\beta)$ of Eq.~(\ref{fermion:eqn:beta}) gives for each $\beta$ a mass function, for which the first term in the OPE, Eq.~(\ref{ope-simple}) vanishes, and so is controlled entirely by the condensate term. However,
as with the analogous NJL model, we have an infinite set of ambiguous definitions of the quark condensate, one for each value of $\beta$, each of which agrees with the chiral condensate in the limit $m_q\rightarrow 0$. This ambiguity seen in Fig.~\ref{cond:fig:quark-beta} arises because, although each solution exhibits the same leading logarithmic behaviour in the ultraviolet limit, the condensates for each are not equal in magnitude, {\it cf.} Eq.~(\ref{ope-simple}). Indeed, the solutions $M^-$, $M^W$ have negative condensates, but we cannot directly use combinations of these mass functions to form a well-defined and unique condensate that coincides with the true condensate contained within $M^+(p^2)$.

%\newpage
\section{Extracting the Condensate}

At very large momenta the tail of the mass function is described by the operator product expansion of Eq.~(\ref{ope-simple}). For QCD, let us introduce the appropriate anomalous dimension factors explicitly, so that
\begin{eqnarray}\label{cond:eqn:opefit}
\vspace{0.5mm}
	M(p^2)_{asym} &=&  {\overline m}\left[ \log\left( p^2/\Lambda_1^{\,2} \right) \right]^{-\gamma_m}\nonumber\\[2.mm]
  &+&\frac{2\pi^2}{3}\frac{\gamma_m {\cal C}}{p^2}\left[ \frac{1}{2}\log\left( p^2/\Lambda_2^{\,2} \right) \right]^{\gamma_m-1} .
\vspace{0.5mm}
\end{eqnarray}
%\begin{eqnarray}\label{cond:eqn:opefit}
%\vspace{0.5mm}
%	M(p^2)_{asym} &=&  {\overline m}\left( \alpha(p_1^2) \right)^{\gamma_m}\nonumber\\[2.mm]
%  &+&\frac{2\pi^2}{3}\frac{\cal C}{p^2}\left( \frac{1}{2}\alpha(p_2^2) \right)^{1-\gamma_m}\;.
%\vspace{0.5mm}
%\end{eqnarray}
where ${\overline m}$ is related to the quantity $m_R(\mu)$ via some renormalisation factors. This provides an excellent representation of all our solutions.
If we included the expression to all orders then the scales $\Lambda_1$,
and $\Lambda_2$ would both be equal to $\Lambda_{QCD}$. However, the leading order forms in Eq.~(\ref{cond:eqn:opefit}) absorb different higher order contributions into the two terms and so $\Lambda_1$ and $\Lambda_2$ are in practice different, as we will discuss below. For large masses the condensate piece, $\cC$, is irrelevant and so it is the leading term that describes the mass function well. In contrast in the chiral limit, ${\overline m}=0$ and so the second term of the OPE describes the behaviour of the mass function. This then accurately determines the scale $\Lambda_2$. Indeed, its value is equal to $\Lambda_{QCD}$. We can then easily extract the renormalisation point independent condensate, $\cC\,\equiv\,-\,\langle \qq \rangle$, from the asymptotics --- see Fig.~5 for the chiral limit.

In this latter case, strictly in the chiral limit, we may also extract the condensate via:
\begin{equation}\label{cond:eqn:tracecondensate}
\vspace{0.5mm}
	-\left<{\overline q}q\right>_\mu = Z_2\left(\mu,\Lambda\right) Z_m\left( \mu,\Lambda \right)N_c\, \tr_D \int^\Lambda \frac{d^4k}{\left( 2\pi \right)^4}\,S\left(k,\mu\right)\, ,
\vspace{0.5mm}
\end{equation}
where $\left<{\overline q}q\right>_\mu$ is the renormalisation dependent quark condensate. At one-loop, this is related to the renormalisation point independent quark condensate:
\begin{equation}
\vspace{0.5mm}
	\left<{\overline q}q\right>_\mu = \left( \frac{1}{2}\log\frac{\mu^2}{\Lambda^2_{QCD}} \right)^{\gamma_m}\left<{\overline q}q\right>\;.
\vspace{0.5mm}
\end{equation}
which we compare with the asymptotic extraction to good agreement.
\begin{figure}[b]
	\centering
	%{\includegraphics*[width=0.90\columnwidth]{figures/condensate-lm}}
	{\includegraphics*[width=0.95\columnwidth]{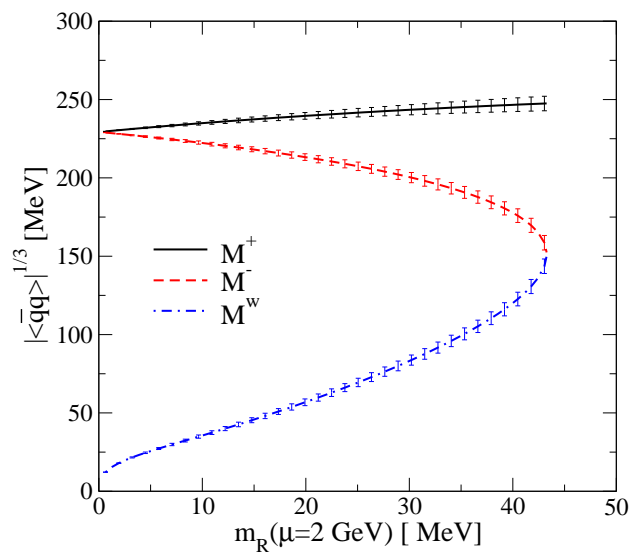}}
	\caption{Condensate extracted through simultaneous fitting of the three solutions to the fermion mass-function in the Maris-Tandy model with $N_f=0$ and $\omega=0.4$ GeV as functions of the current quark mass defined at 2 GeV in a
modified momentum subtraction scheme.}\label{cond:fig:cond-lm}
\end{figure}

%\noindent
However, for small quark masses, where the condensate is believed to play a sizeable role, we cannot apply Eq.~(\ref{cond:eqn:tracecondensate}), since it acquires a quadratic divergence, {\it cf.} Eq.~(\ref{ope-simple}). Indeed, it is the elimination of this that inspired the original Eq.~(\ref{cond:eqn:cond-roberts}) and later Eq.~(\ref{cond:eqn:cond-beta}), which we have seen lead to a wholly ambiguous definition of the physical condensate. Nevertheless, one can  attempt to fit both terms of the OPE in Eq.~(\ref{cond:eqn:opefit}) to the tail of the mass function, $M^+$ for instance. While a value for the condensate can then be extracted, this procedure is not at all reliable because of the difficulty in resolving the two functions in the OPE from one another and in fixing the appropriate scales, $\Lambda_1$ and $\Lambda_2$.
%\end{figure}
%\vspace{5mm}
%%\begin{figure}[ht]
%	\centering
%	{\includegraphics*[width=0.90\columnwidth]{figures/condensate-lm-compare}}
%	\caption{Combinations of true condensate formed to compare with Fig.~\ref{cond:fig:quark-beta}, shown as functions of the current quark mass defined at 1 GeV.}\label{cond:fig:cond-lm-comp}

It is to this last point which we now turn. Instead of one single solution, we now have three solutions to the same model, each with identical running of the current-quark mass (the first term in Eq.~(\ref{cond:eqn:opefit})) in the ultraviolet region and differing only by their values of the condensate. Thus it is possible to fit Eq.~(\ref{cond:eqn:opefit}) simultaneously to the three mass functions $M^\pm$, $M^W$. The scales $\Lambda_1$ and $\Lambda_2$ are determined separately for each value of the current mass. Remarkably,  within the given model, 
they are exactly the same for the range of quark masses we consider
with $\Lambda_1\sim 2\Lambda_{QCD}$ and $\Lambda_2\sim \Lambda_{QCD}$ respectively. The condensates $\cC^\pm$ and $\cC^W$ are then determined
in an accurate and stable way. 
For this to work the solutions have to be found to an accuracy of 1 part in $10^{8}$. This fitting is performed using a modified Levenberg-Marquardt algorithm with appropriate weights added to give better behaviour at large momentum in accord with perturbation theory. The results for the phenomenological model employed here are given in Fig.~\ref{cond:fig:cond-lm}. The error bars reflect the accuracy
with which the mass functions, representable by two terms in the OPE expression, Eq.~(\ref{cond:eqn:opefit}), are separable with the anomalous dimensions specified. In contrast to the condensate defined by Eq.~(\ref{cond:eqn:cond-roberts}), we find that in the limited mass range investigated, the condensate increases as a function of $m_q$. At the critical point $m_{cr}(\mu=2$~GeV)=44 MeV, we find the ratio for the condensate to the chiral limit with $N_f=0$ to be (Fig.~\ref{cond:fig:cond-lm}):
%\begin{figure}[ht]
%\centering
%      {\includegraphics*[width=0.90\columnwidth]{figures/cond-mt-0-4}}
%	\caption{Condensate for Maris-Tandy Model with $N_f=0$, $\omega=0.4$ GeV}\label{cond:mt-0-4b}
%\end{figure}
%\vspace{5mm}
\begin{figure}[t]
  \centering
      %{\includegraphics*[width=0.90\columnwidth]{figures/cond-mt-4-5}}
	%\caption{Condensate for Maris-Tandy Model with $N_f=4$, $\omega=0.5$ GeV as a function of current quark mass defined at 1~GeV, to be compared with Fig.~\ref{cond:fig:cond-lm}.}\label{cond:mt-4-5b}\label{cond:mt-4-5b}
      {\includegraphics*[width=0.95\columnwidth]{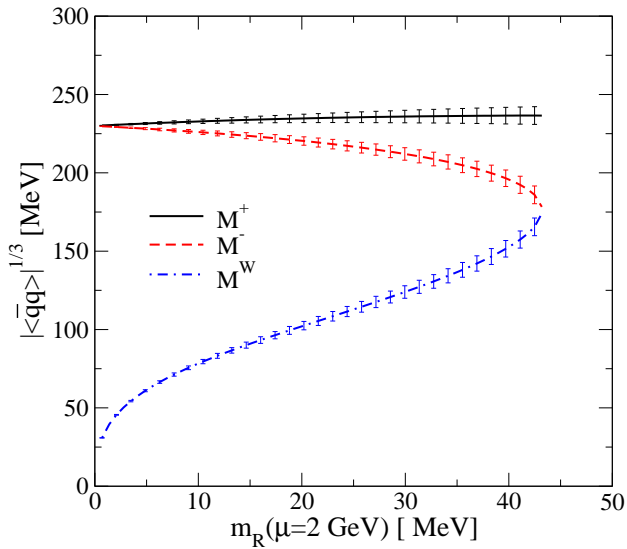}}
	\caption{Condensate for Maris-Tandy Model with $N_f=4$, $\omega=0.4$ GeV as a function of current quark mass defined at 2~GeV, to be compared with Fig.~\ref{cond:fig:cond-lm}.}\label{cond:mt-4-4b}
\end{figure}
\begin{equation}
%	\frac{\left<{\overline q}q\right>_{m=50{\rm MeV}}}{\left<{\overline q}q\right>_{m=0}} = 1.29
	\left<{\overline q}q\right>_{m=50\,{\rm MeV}}/\left<{\overline q}q\right>_{m=0} = 1.24
\end{equation}
To estimate the errors in this determination, we can form combinations of these condensates in the same way as defined in Eq.~(\ref{cond:eqn:cond-beta}), favourably reproducing the same results of Fig.~\ref{cond:fig:quark-beta}.
In Fig.~\ref{cond:mt-4-4b} is a similar plot with $N_f=4$ and $\omega=0.4$~GeV, illustrating how $m_{cr}$ changes compared with Fig.~\ref{cond:fig:cond-lm}.

We see that within errors the condensate is found to increase with quark mass.
This rise at small masses was anticipated by Novikov {\it et al.}~\cite{Novikov:1981xj} combining a perturbative chiral expansion with QCD sum-rule arguments. That the chiral logs relevant at very small $m_q$ are barely seen is due to the quenching of the gluon and the rainbow approximation of Eq.~(\ref{cond:eqn:quarksde1}). As we will show in the next section when we model more complex interactions, including matching with the lattice, this effect remains small.
%\newpage
\section{More sophisticated  models of QCD interaction}\label{cond:sect:sophistry}
We now consider the consequences of using more sophisticated
vertex structure for the quark-gluon interaction in the
quark Dyson-Schwinger equation, Fig.~\ref{quark:sde}.
The first framework we study is a truncation scheme
introduced in \cite{Fischer:2003rp,Fischer:2003zc}. It
involves replacing the bare quark-gluon vertex of  Sect.~\ref{cond:sect:maris} with the
Curtis-Pennington (CP) vertex~\cite{Curtis:1990zs}, thus
ensuring multiplicative renormalizability for the fermion
propagator. In the Yang-Mills sector of QCD ans\"atze for
ghost and gluon interactions have been introduced, which
enable a self-consistent solution for the ghost and gluon
propagators.
The second scheme we shall investigate is an ansatz for the
quark-gluon vertex, which has been fitted to lattice results,
and was previously employed in Ref~\cite{Fischer:2005nf,torus:update}.

\section*{Continuum studies: CP vertex}\label{cond:sect:CP}
\begin{figure}[b]
\centering
{\includegraphics*[width=\columnwidth]{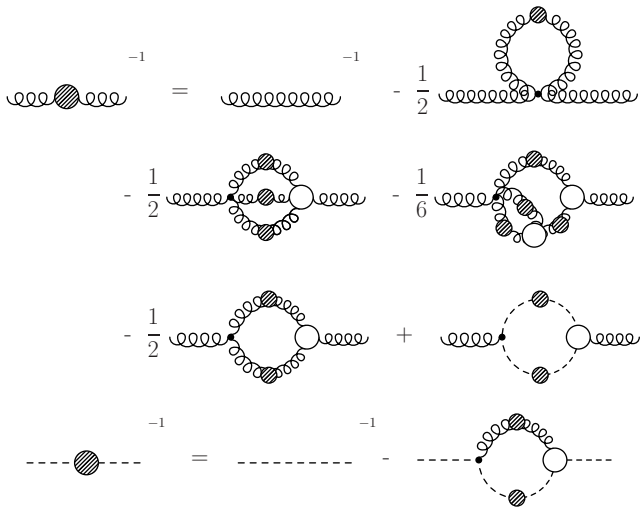}}
 \caption{Dyson-Schwinger equations for the gluon and ghost
          propagator. Filled circles denote dressed propagators
      and empty circles denote dressed vertex functions.}
      \label{fig:ghostglue}
\end{figure}

\noindent
In this truncation scheme we use explicit solutions for the
Dyson-Schwinger equations for the ghost and gluon propagators,
given diagrammatically in Fig.~\ref{fig:ghostglue}.
This system of equations has been solved numerically
in \cite{Fischer:2002hn}. The relevant ans\"atze for
the ghost-gluon and triple-gluon vertices have been discussed
in the literature \cite{Fischer:2002hn,Fischer:2003rp,Fischer:2006ub}.
The solutions for the ghost and gluon propagator
\begin{eqnarray}
D_{G}(p)       &=& - \frac {G(p^2)}{p^2} \;, \label{Gh-prop}\\
D_{\mu \nu}(p) &=& \left(\delta_{\mu \nu} - \frac{p_\mu p_\nu}{p^2} \right)
                  \frac{Z(p^2)}{p^2} \;, \label{Gl-prop}
\end{eqnarray}
can be represented accurately by
\begin{eqnarray}
Z(p^2) &=& \left( \frac{\alpha(p^2)}{\alpha(\mu)} \right)^{1+2\delta}
R^2(p^2) \,, \label{glue}\\
G(p^2) &=& \left( \frac{\alpha(p^2)}{\alpha(\mu)} \right)^{-\delta}\quad
R^{-1}(p^2)  \,,\\\nonumber
&&\hspace{-64pt}{\rm with}\\ 
R(p^2) &=& \frac{c \,(p^2/\Lambda^2_{YM})^{\kappa}+d
\,(p^2/\Lambda^2_{YM})^{2\kappa}}
{1+ c \,(p^2/\Lambda^2_{YM})^{\kappa}+d \,(p^2/\Lambda^2_{YM})^{2\kappa}}
\,, \nonumber\\
\label{ghost}
\end{eqnarray}
with the scale $\Lambda_{YM}=0.658 \,\mbox{GeV}$, the coupling
$\alpha(\mu)=0.97$ and the parameters $c=1.269$ and $d=2.105$ in the
auxiliary function $R(p^2)$. The quenched anomalous dimension $\gamma$
of the gluon is related to the anomalous dimension $\delta$ of the
ghost by $\gamma=-1-2\delta$ and $\delta=-9/44$ for $N_f=0$. The
infrared exponent $\kappa = (93-\sqrt{1201})/98
\approx 0.595$ \cite{Lerche:2002ep}. The running coupling $\alpha(p^2)$
is defined via the nonperturbative ghost-gluon vertex,
\begin{equation}
\alpha(p^2) = \alpha(\mu)\, G^2(p^2)\, Z(p^2)
\end{equation}
and can be represented by
\begin{eqnarray}
\alpha(p^2) &=& \frac{1}{1+p^2/\Lambda^2_{YM}}
\bigg[\alpha(0) + p^2/\Lambda^2_{YM} \times \nonumber\\
&& \hspace*{-2mm}\left.\frac{4 \pi}{\beta_0}
\left(\frac{1}{\ln(p^2/\Lambda^2_{YM})}
- \frac{1}{p^2/\Lambda_{YM}^2 -1}\right) \right]\,. \end{eqnarray}
The value $\alpha(0) \approx 8.915/N_c$ is known from
an analytical infrared analysis \cite{Lerche:2002ep}.
\begin{figure}[b]
\centering
%{\includegraphics*[width=0.88\columnwidth]{figures/cond-cp}}
     {\includegraphics*[width=0.95\columnwidth]{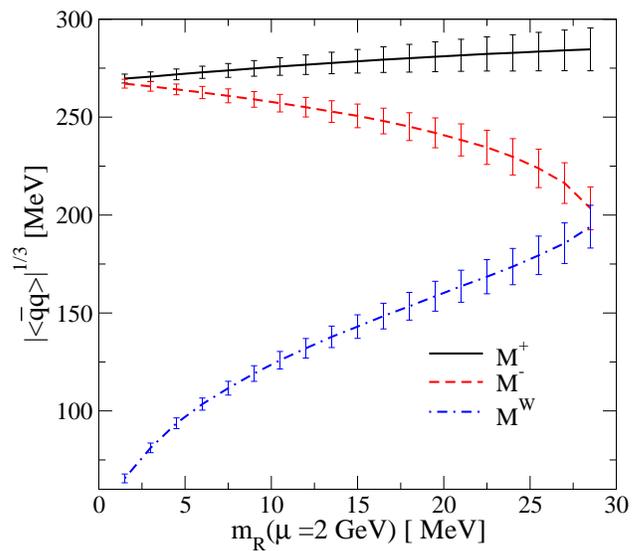}}
   \caption{Condensate for CP-Vertex Model. $N_f=0$}\label{cond:cp}
\end{figure}

In the quark-DSE Eq.~(\ref{cond:eqn:quarksde1}) we use the solution of
Eq.~(\ref{glue}) for the gluon propagator together with an ansatz of the
form
\begin{equation}
\Gamma_\nu(q,k) = V_\nu^{Abel}(p,q,k) \, W^{\neg Abel}(p,q,k),
\label{cpvertex}
\end{equation}
where $p$ and $q$ denote the quark momenta and $k$ the gluon momentum.
The ansatz factorises into an Abelian and an non-Abelian part which are
specified and discussed in detail in \cite{Fischer:2003rp}. Here we
only need to remark that the Abelian part $V_\nu^{abel}$ is identical to
the CP vertex \cite{Curtis:1990zs}. This construction
carries further tensor structure in addition to the $\gamma_\mu$-piece,
which makes it an interesting ansatz in comparison with the simple model
in  Sect.~\ref{cond:sect:maris}. The corresponding numerical solutions for the
quark propagator are discussed in detail in \cite{Fischer:2003rp}.
Here we are only interested in the chiral condensate as a function of
the current quark mass. The corresponding results can be found
in Fig.~\ref{cond:cp}. Despite the complicated tensor structure
of the vertex Eq.~(\ref{cpvertex}) we find similar results for the
condensate as previously.
We were able to extract the condensate from all three solutions
$M^\pm$ and $M^W$, again for a restricted region of $m_q < m_{cr}$. The physical condensate rises again slightly
for small current quark masses and bends down for larger ones. The
critical value of $m_{cr}$ is found to be $20$ MeV at $\mu=19$ GeV.
This corresponds to $30$ MeV at $\mu=2$ GeV.
With the parameters of \cite{Fischer:2003rp}, the condensate in the chiral limit is 270 MeV, rather than the phenomenological 235 MeV we have used. However, its dependence with quark mass hardly depends on this exact value and so
at the critical point we obtain the ratio
\begin{equation}
   \left<{\overline q}q\right>_{m=30\,{\rm MeV}}/\left<{\overline
q}q\right>_{m=0}\; =\; 1.175\quad . 
\end{equation}
As with the phenomenological model considered in  Sect.~\ref{cond:sect:maris} we
find a considerable increase of the chiral condensate with the current quark
mass.

\section*{Lattice Model}\label{cond:sect:lattice}
The third model we investigate has been defined in \cite{Fischer:2005nf}.
The idea is to solve the coupled system of gluon, ghost and quark
Dyson-Schwinger equations on a compact manifold with periodic boundary
conditions, similar to lattice QCD. For the vertices in the Yang-Mills
sector
the same truncation scheme as in the last section is employed. However,
for the quark-gluon vertex an ansatz has been specified such that lattice
results for the quark propagator have been reproduced on a similar manifold.
Solving the system also in the infinite volume/continuum limit one can then
study volume effects in the pattern of dynamical chiral symmetry
breaking \cite{Fischer:2005nf,torus:update}.

In the infinite volume/continuum limit, i.e on  $\mathbb{R}^4$, the
solutions
for the ghost and gluon propagator are given by Eq.~(\ref{glue}) and
Eq.~(\ref{ghost}).
The ansatz for the quark-gluon vertex is
\begin{equation}
\Gamma_\nu(k,\mu^2) \;=\; \gamma_\nu\, \Gamma_{1}(k^2)\,
\Gamma_{2}(k^2,\mu^2)\, \Gamma_{3}(k^2,\mu^2) \label{v1}
\end{equation}
with the components
\begin{figure}[t]
   \centering
%    \vspace{5mm}
     %{\includegraphics*[width=0.88\columnwidth]{figures/cond-lat}}
     {\includegraphics*[width=0.98\columnwidth]{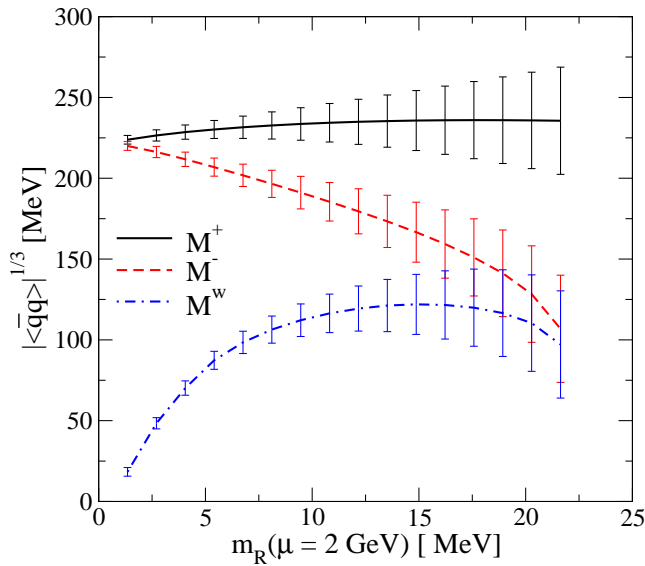}}
   \caption{Condensate for Lattice Model. $N_f=0$}\label{cond:lat}
\end{figure}

\begin{table}[b]
\begin{center}
\begin{tabular}{|c||c|c|c|c|c|c|}
\hline            & \hspace*{2mm} h               \hspace*{2mm}
     & \hspace*{1mm} $\Lambda_g$     \hspace*{1mm}
     & \hspace*{1mm} $\Lambda_{QCD}$ \hspace*{1mm}
     & \hspace*{2mm} $a_1$           \hspace*{2mm}
     & \hspace*{2mm} $a_2$           \hspace*{2mm}
     & \hspace*{2mm} $a_3$           \hspace*{2mm}\rule[-1mm]{0mm}{5mm}\\
         & & \hspace*{1mm} (GeV) \hspace*{1mm}
     & \hspace*{1mm} (GeV) \hspace*{1mm}  & & & \\
     \hline\hline
overlap   &  1.31 & 1.50  & 0.35 & 25.58 & 3.44  &  2.23
\rule[-3mm]{0mm}{7mm} \\\hline
\end{tabular}
\caption{Parameters used in the vertex model,
Eqs.~(\ref{v1}-\ref{v4}).\label{tab2}}
\end{center}
\end{table}
\begin{eqnarray}
\Gamma_{1}(k^2) &=&  \frac{\pi \gamma_m}{\ln(k^2/\Lambda_{QCD}^2 +\tau)}\,,
\label{v2}\\[3mm]
\Gamma_{2}(k^2,\mu^2) &=& G(k^2,\mu^2)\ G(\zeta^2,\mu^2)\
\widetilde{Z}_3(\mu^2) \nonumber\\[-2.mm]
&& \\[-2.mm]
&&\hspace{10mm} \times \ h \ [\ln(k^2/\Lambda_{g}^2 +\tau)]^{1+\delta}
\label{v3}\nonumber\\[3mm]
\Gamma_{3}(k^2,\mu^2) &=& Z_2(\mu^2)\;
\frac{a(M)+k^2/\Lambda_{QCD}^2}{1+k^2/\Lambda_{QCD}^2}\,, \label{v4}
\end{eqnarray}
where $\delta=-9/44$ is the (quenched) one-loop anomalous dimension of
the ghost, $\gamma_m=12/33$ the corresponding anomalous dimension of
the quark and $\tau = e-1$ acts as a convenient infrared cutoff for
the logarithms. The quark mass dependence of the vertex is parametrised by
\begin{equation}
a(M)\; =\; \frac{a_1}{1 + a_2 M(\zeta^2)/\Lambda_{QCD} + a_3
M^2(\zeta^2)/\Lambda_{QCD}^2},
\label{am}
\end{equation}
where $M(\zeta^2)$ is determined during the iteration process at $\zeta=2.9$
GeV.
The parameters have been fitted to lattice results using
a staggered \cite{Bowman:2002bm} and an overlap action \cite{Zhang:2004gv}.
Here we use only the fit to the overlap quark; the corresponding parameters
are 
given in table \ref{tab2}. 
%\begin{eqnarray}
%\nonumber
%h&=&1.31\quad ,
%\Lambda_g\;=\;1.50\,{\rm GeV}\;,\qquad \Lambda_{QCD}\;=\;0.35\,{\rm GeV}\;,\\\nonumber
%a_1&=&25.58\;,\quad a_2\;=\;3.44\;,\quad a_3\;=\;2.23\; .
%\end{eqnarray}
Explicit solutions for the quark propagators
are discussed in \cite{Fischer:2005nf}.

This modelling provides a link between lattice results and the continuum.
The vertex function, as defined by Eqs.~(35-37), which is currently fitted to the lattice data, involves several different scales: $\Lambda_g$, $\Lambda_{YM}$ and $\Lambda_{QCD}$. In the continuum this results in a failure to reproduce the perturbatively determined anomalous dimensions in the OPE, Eq.~(\ref{cond:eqn:opefit}). The extraction of the tiny condensate term in the OPE is very sensitive to these anomalous dimensions. Combined with the effect the uncertainties in determining the multiple scales has on $\Lambda_1$ and $\Lambda_2$ in Eq.~(\ref{cond:eqn:opefit}) leads here to much larger errors than in our previously modellings. Nevertheless, for the massive condensate we again find solutions similar to the previous
sections.
The condensate corresponding to the $M^{\pm}$ solutions are given in
Fig.~\ref{cond:lat}
and the critical value $m_{cr}$ is $22$ MeV, see
Table~\ref{table:cplatticecritical}.
The large errors point to the need for further studies of matching lattice on a torus to the continuum, if we are to extract reliable infinite volume, continuum quantities like the quark condensate. That is for the future.
The ratio of the condensates is here
   $\left<{\overline q}q\right>_{m=22\,{\rm MeV}}/\left<{\overline
q}q\right>_{m=0}\;=\;1.17$.
This is a little lower than for the previous models of Eqs.~(24,33).
However, uncertainties in extraction are considerably larger.
Nevertheless the ratio is still bigger than one.

%%%%%%%%%%%%
%For our two models with the more complicated vertex structure, or fitted to
%%%%%lattice data:
\begin{table}[h]
\begin{center}

\begin{tabular}{|c||ccr|}
%    $N_f$ & Model & \hspace{2.5mm}CP\hspace{2.5mm} & Lattice &
%\\[1mm]
\hline
%&&&&\\[-3.5mm]
%    0 & \rule{0ex}{2.5ex} $m_{cr}(\zeta=19$~GeV) & 20 & 16 &[MeV]\\[1.mm]
%    0 & \rule{0ex}{2.5ex} $m_{cr}(\zeta=1$~~GeV) & 30 & 24 &[MeV]\\[1.mm]
   Model & \hspace{2.5mm}CP\hspace{2.5mm} & Lattice & \\[1mm]\hline\hline
&&&\\[-3.5mm]
\rule{0ex}{2.5ex} $m_{cr}(\mu=19$~GeV) & 20 & 16 &[MeV]~~\\[1.mm]
\rule{0ex}{2.5ex} $m_{cr}(\mu=2$~~GeV) & 30 & 22 &[MeV]~~\\[1.mm]
\hline
\end{tabular}
\caption{The critical mass for our quenched CP and Lattice
model.}\label{table:cplatticecritical}
\end{center}
\end{table}

\section{Condensate beyond the critical mass range: Noded Solutions}

We see in Figs.~\ref{cond:fig:cond-lm}, \ref{cond:mt-4-4b} that the $M^-$ and $M^W$ solutions bifurcate below $m_{cr} \simeq 43.4 (44.0)$ MeV with $\omega =0.4$ GeV for $N_f=0(4)$ respectively.
 But what about the value of the condensate for the physical solution $M^+$ beyond the region where $M^-$ and $M^W$ exist, {\it i.e.} $m_R(\mu) > m_{cr}$?
Having accurately determined the scales $\Lambda_1$ and $\Lambda_2$ in the OPE of Eq.~(\ref{cond:eqn:opefit}) in the region where all 3 solutions exist, we
could
 just continue to use the same values in fitting the physical $M^+$ solution alone and find its condensate.
Unfortunately, this would make
it difficult to produce realistic errors as the quark mass increases.
\begin{figure}[t]
\centering      %{\includegraphics*[width=0.88\columnwidth]{lanlfig6.eps}}
     {\includegraphics*[width=0.95\columnwidth]{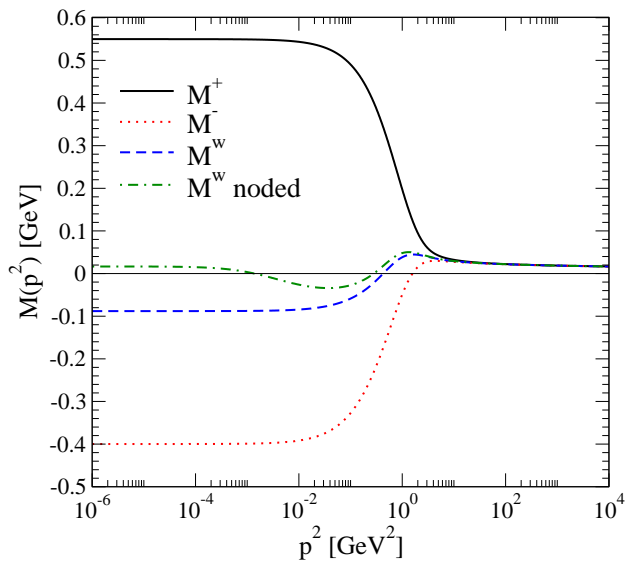}}
   \caption{Momentum dependence of the 4 solutions for the fermion
mass-function in the Maris-Tandy model with $m=20$ MeV at $\mu=19$ GeV,
$N_f$=4, $\omega=0.4$ GeV.}\label{cond:noded}
\end{figure}
%\vspace{5mm}

%\begin{figure}[p]
%\centering      
%     {\includegraphics*[width=0.62\columnwidth]{lanlfig6}}
%   \caption{Momentum dependence of the 4 solutions for the fermion
%mass-function in the Maris-Tandy model with $m=20$ MeV at $\mu=19$ GeV,
%$n_F$=4, $\omega=0.4$ GeV.}\label{cond:noded}
%\vspace{4mm}
%\centering     
%     {\includegraphics*[width=0.62\columnwidth]{lanlfig7}}
%   \caption{Current quark mass dependence of the condensates for Maris-Tandy
%model with $n_F=4$, $\omega=0.4$ GeV, including the noded solution of
%Fig.~\ref{cond:noded}.}\label{cond:nodecond}
%\end{figure}

However,
as soon as one allows for solutions for the fermion mass-function that are not positive definite, one exposes a whole series of variants on the solutions $M^-$, $M^W$ we have already considered. Thus there are {\it noded} solutions, which have also been discovered recently in the context of a simple Yukawa theory  by Martin and Llanes-Estrada~\cite{Martin:2006qd}. 
These noded solutions are only accessible if sufficient numerical precision is used. For instance, representing the dressing functions by a Chebyshev expansion does not provide the required accuracy with sufficient smoothness to reproduce the OPE form of Eq.~(21). We illustrate this within the Maris-Tandy model, for instance with $N_f=4$ and $\omega=0.4$, in Fig.~\ref{cond:noded}. There the four solutions we have found are displayed.

\begin{figure}[b]
   \centering
     %{\includegraphics*[width=\columnwidth]{figures/cond-mt-4-4}}
     {\includegraphics*[width=0.98\columnwidth]{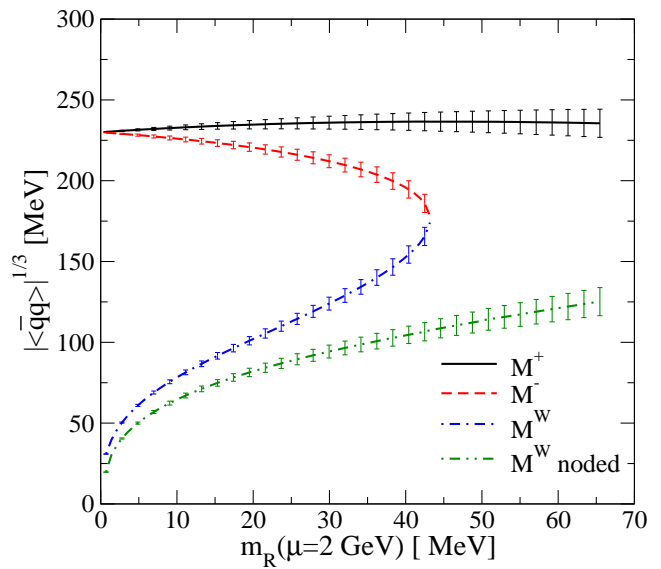}}
   \caption{Current quark mass dependence of the condensates for Maris-Tandy
model with $N_f=4$, $\omega=0.4$ GeV, including the noded solution of
Fig.~\ref{cond:noded}.}\label{cond:nodecond}
\end{figure}

It is interesting to note that this noded solution is not limited to the same domain that restricts $M^-$ and $M^W$. These noded solutions do develop a singularity in $M(p^2)$ beyond $m=51.4$ MeV at $\mu=2$ GeV. However, this is compensated for by a zero in $\cA(p^2)$, Eq.~(\ref{fermion:eqn:AB}), until $m=66.3$ MeV.  Thus there exists a solution with a well-defined ultraviolet running of the quark mass exactly as the $M^+$ solution, as far as $m=66.3$ MeV. While at small quark masses we have all four solutions, at larger masses there are still two. Consequently, we can confirm that 
the scales $\Lambda_1$ and $\Lambda_2$ of Eq.~(\ref{cond:eqn:opefit}) as fixed for  $m < m_{cr}$ are still well-determined by our fit procedure for
$m > m_{cr}$ and so deduce the condensates.
Indeed, fitting the $M^+$ and $M^W_{noded}$ at each value of $m_R(\mu)$ with common scales in the OPE equation, Eq.~(\ref{cond:eqn:opefit}) allows the condensate for the physical solution to be found for much larger quark masses, as shown in Fig.~\ref{cond:nodecond}. Indeed, these fits confirm that $\Lambda_1$ and $\Lambda_2$ are independent of $m_R(\mu)$. We can then fit the remaining $M^+$ solutions shown in Fig.~\ref{masses} to give the physical condensate shown in Fig.~\ref{cond:nodecond} for acceptable values of $\omega$ as determined by ~\cite{Alkofer:2002bp}.

\begin{figure}[t]
  \centering
      {\includegraphics*[width=0.98\columnwidth]{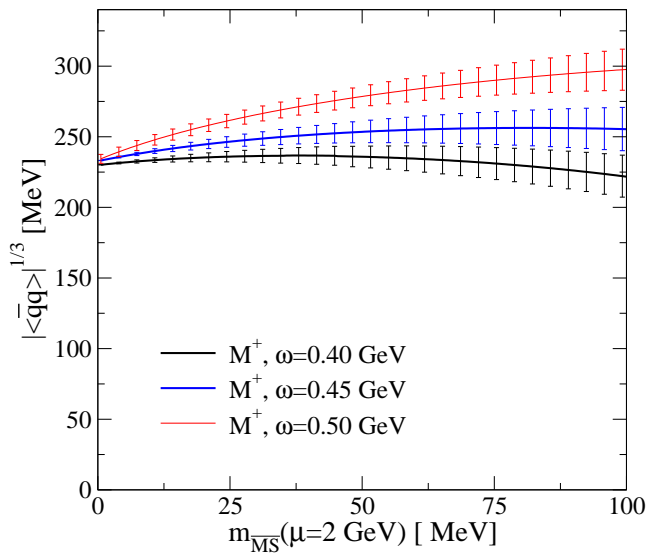}}
	\caption{Condensate for Maris-Tandy Model with $N_f=4$, $\omega=0.4,0.45,0.5$ GeV as a function of current quark mass defined at 2~GeV in $\overline{MS}$ scheme.}\label{cond:m45MS}
\end{figure}

 In Fig.~\ref{cond:m45MS} we scale the quark mass from $\mu = 2$ GeV in the (quark-gluon) MOM scheme by one loop running to the
$\overline{MS}$ scheme at 2 GeV using the relationship between $\Lambda_{MOM}$ and $\Lambda_{\overline{MS}}$ for 4 flavours deduced by Celmaster and Gonsalves~\cite{Celmaster:1979km}. In this latter scheme the strange quark mass is $\sim 95$~MeV as given in the PDG Tables~\cite{PDG}.
Within the range of the Maris-Tandy modelling of strong coupling QCD, we find the ratio of the condensates at the strange quark mass to the chiral limit is
\begin{equation}
	\left<\qq\right>^{1/3}_{m(\overline{MS})\,=95\,{\rm MeV}}/\left<\qq\right>^{1/3}_{m=0}\;=\;(\,1.1\,\pm\,0.2\,)\; .
\end{equation}
in a world with 4 independent flavours. 
Moreover, here all the quarks have the same mass and there is no mixing between  different hidden flavour pairs. Elsewhere we will illustrate the change that occurs in solving the quark Schwinger-Dyson equations with 2 flavours
of very small mass $m_{u,d}$ and 1 flavour with variable mass. Of course, in the quenched case quark loops decouple and exactly replicate the results given here.  

%Chang {\it et al.} have proposed that the fact that the $M^-$ and $M^W$ solutions only exist for some domain of quark masses, $m_q\,\in\,[0,m_{cr}]$, is directly linked to the domain of convergence of the chiral expansion for $M^+$~\cite{Hatsuda:1990tt,Meissner:1994wy}. While the existence of multiple solutions for the fermion mass function is essential for the extraction of the condensate, we note that distinct domains exist for
%the different solutions and that the simplest noded solution exists in a larger domain. Though the value of $m_{cr}$ is indeed some measure of the
%range of validity of the chiral expansion, the fact that $m_{cr}$
%is both strongly model and solution dependent indicates its value is no more than a guide and not likely to be the exact bound claimed by Chang {\it et al.} 

What we have shown here is that there is a robust method of determining the value of the $\qq$ condensate beyond the chiral limit based on the Operator Product Expansion. Of course, as the
quark mass increases the contribution of the condensate to the behaviour of the mass function, Fig.~\ref{masses}, becomes relatively less important and so the errors on the extraction of the physical condensate increases considerably.
Nevertheless, the method is reliable up to and beyond the strange quark mass.
Alternative definitions are not.

%As soon as one allows for solutions for the fermion mass-function that are not positive definite, one exposes a whole series of variants on the solutions $M^-$, $M^W$ we have already considered. Thus there are {\it noded} solutions, which have also been discovered recently by Martin~\emph{et~al}~\cite{Martin:2006qd} in the context of a simple Yukawa theory. We illustrate this within the Maris-Tandy model, for instance with $N_f=4$ and $\omega=0.4$, in Fig.~\ref{cond:noded}, where the four solutions we have found are displayed. Again the fourth {\it noded} solution has the same running current mass behaviour as the other three solutions, and so again a \lq\lq condensate'' (in the sense of the OPE of Eq.~(\ref{cond:eqn:opefit})) can be extracted. How its behaviour with quark mass compares with the other condensates in the same model is shown in Fig.~\ref{cond:nodecond}. It is interesting to note that this noded solution is not limited to the same domain that restricts $M^-$ and $M^W$. Despite these noded solutions developing a singularity beyond $m=36.4$ MeV at $\mu=19$ GeV, they continue to exist until $m=47$ MeV with a well-defined ultraviolet running of the quark mass. This latter critical point corresponds to a $m'_{cr}=79$ MeV at $\mu=1$ GeV, which we compare with the $m_{cr}=52$ MeV of Table~\ref{table:mariscrit}.

While the existence of multiple solutions to the fermion Schwinger-Dyson equation is essential for our method, only the $M^+$ solution has any physical significance and the others are mathematical curiosities. Those, like $M^-$, $M^W$ and $M^W_{noded}$, only exist in restricted domains. In contrast, the physical solution exists for all current quark masses, even if we cannot reliably extract the value of the corresponding condensate from the OPE. While it is clear that the radius of convergence of the chiral expansion  of $M^+$~\cite{Hatsuda:1990tt,Meissner:1994wy} in terms of the quark mass  has a scale of order of $\Lambda_{QCD}$ or equally $\left(-\langle {\overline q}q \rangle\right)^{1/3}$,
 is this scale set by $m_{cr}$ of Eq.~(5)? Chang {\it et al.} claim it is~\cite{Chang:2006bm}. However, the bifurcation point for the unphysical solutions differs whether they are noded and not, cf. Figs.~9 and 14, and in turn each is highly model-dependent.
This  makes it difficult to claim that the value of $m_{cr}$ of Eq.~(5) is the key parameter of the radius of the convergence of the chiral expansion for $M^+$.

%\section*{Improvements}
%\begin{itemize}
%	\item The fit can be further constrained by fitting to the ``mixed condensate'' contained within the differences $(M^+ - M^-)$, $(M^+-M^W)$ and $(M^--M^W)$.
%	\item Obtaining the Wigner mode allows for an additional check on the accuracy of the extraction procedure.
%\end{itemize}

\section{Summary}
Within the NJL model and the Schwinger-Dyson approach to QCD, we have investigated the three inequivalent solutions, called $M^+$, $M^-$ and $M^W$, to the mass gap equation that exist within the interval $\cD(m) = \left\{ m : 0 \le m \le m_{cr}\right\}$. By ensuring each were solved using the same renormalisation conditions, we found that  each solution exhibited the same running of the current-quark mass in the ultraviolet: differentiated solely by their infrared behaviour and value of the quark condensate.

Though it was not possible to define the condensate unambiguously  by simply taking combinations of $M^+$, $M^-$ and $M^W$, the increased information available on the domain $\cD(m)$ by having three solutions permits a reliable extraction of the condensate through simultaneous fitting of these to the OPE. In
addition, we were able to obtain a fourth (noded) solution. This is only possible if the equations are solved to high numerical accuracy. Though this fourth solution  violates the physical requirement of positivity, it has allowed us to extract the condensate beyond $m_{cr}$ and into the region of the physical strange quark mass.

We have investigated a number of models for the strong coupling (infrared) behaviour of the quark-gluon interaction of QCD and  found in all cases that the condensate, corresponding to the solution with a positive-definite mass function, increases moderately 
with current quark mass in the region under consideration.
This is in contrast to the QCD sum-rule calculations of Refs. 6-8. Typically we find an increase of 30\% from the chiral limit to a current ${\overline{MS}}$ mass of 100 MeV at a scale of 2 GeV. Only at still larger quark masses does the condensate
significantly decrease.

%\vspace{5mm}

\begin{acknowledgments}

RW is grateful to the UK Particle Physics and Astronomy Research Council (PPARC) for the award of a research studentship. CSF acknowledges a Helmholtz Young Investigator award VH-NG-332.
We thank Roman Zwicky and Dominik Nickel for interesting discussions.
This work
was supported in part by the EU RTN
Contracts HPRN-CT-2002-00311, \lq\lq EURIDICE''
and MRTN-CT-2006-035482, \lq\lq FLAVIAnet''. Two of us (CSF and MRP) wish to thank ECT* and the organisers of the Workshop on Quark Confinement in Trento, where this work was completed.
\end{acknowledgments}

\end{document}